\documentstyle[12pt]{article}

\newcommand{\be}{\begin{equation}}
\newcommand{\ee}{\end{equation}}

\newcommand{\bea}{\begin{eqnarray}}
\newcommand{\eea}{\end{eqnarray}}
\newcommand{\ba}{\begin{array}}
\newcommand{\ea}{\end{array}}
\newcommand{\beqa}{\begin{eqnarray}}
\newcommand{\eeqa}{\end{eqnarray}}

\newcommand{\NP}[1]{Nucl. Phys.\ {\bf #1}\ }
\newcommand{\PL}[1]{Phys. Lett.\ {\bf #1}\ }

\newcommand{\PR}[1]{Phys. Rev.\ {\bf #1}\ }
\newcommand{\PRL}[1]{Phys. Rev. Lett.\ {\bf #1}\ }

\newcommand{\h}{{1\over2}}

\newcommand{\del}{\partial}

\newcommand{\Tr}{{\rm Tr}}

\newcommand{\D}{\delta}
\newcommand{\DE}{\Delta}

\newcommand{\ka}{\kappa}

\newcommand{\ie}{{\it i.e. }}

\newcommand{\ssu}{$SU(2)_L\times SU(2)_R\times U(1)_{B-L}\,$}

\newcommand{\sur}{$SU(2)_R$}
\newcommand{\matr}{\left( \begin{array}}
\newcommand{\ematr}{\end{array} \right)}

\newcommand{\lsim}
{{\;\raise0.3ex\hbox{$<$\kern-0.75em\raise-1.1ex\hbox{$\sim$}}\;}}
\newcommand{\gsim}
{{\;\raise0.3ex\hbox{$>$\kern-0.75em\raise-1.1ex\hbox{$\sim$}}\;}}

\catcode`\@=11
\long\def\@makefntext#1{
\protect\noindent \hbox to 3.2pt {\hskip-.9pt
$^{{\ninerm\@thefnmark}}$\hfil}#1\hfill}		%CAN BE USED

 \def\@makefnmark{\hbox to 0pt{$^{\@thefnmark}$\hss}}  %ORIGINAL

\def\ps@myheadings{\let\@mkboth\@gobbletwo
\def\@oddhead{\hbox{}
\rightmark\hfil\ninerm\thepage}
\def\@oddfoot{}\def\@evenhead{\ninerm\thepage\hfil
\leftmark\hbox{}}\def\@evenfoot{}
\def\sectionmark##1{}\def\subsectionmark##1{}}

\newcounter{sectionc}\newcounter{subsectionc}\newcounter{subsubsectionc}
\renewcommand{\section}[1] {\vspace{0.6cm}\addtocounter{sectionc}{1}
\setcounter{subsectionc}{0}\setcounter{subsubsectionc}{0}\noindent
	{\bf\thesectionc. #1}\par\vspace{0.4cm}}
\renewcommand{\subsection}[1] {\vspace{0.6cm}\addtocounter{subsectionc}{1}
	\setcounter{subsubsectionc}{0}\noindent
	{\it\thesectionc.\thesubsectionc. #1}\par\vspace{0.4cm}}
\renewcommand{\subsubsection}[1]
{\vspace{0.6cm}\addtocounter{subsubsectionc}{1}
	\noindent {\rm\thesectionc.\thesubsectionc.\thesubsubsectionc.
	#1}\par\vspace{0.4cm}}

\newcounter{appendixc}
\newcounter{subappendixc}[appendixc]
\newcounter{subsubappendixc}[subappendixc]

\renewcommand{\appendix}[1] {\vspace{0.6cm}
        \refstepcounter{appendixc}
        \setcounter{figure}{0}
        \setcounter{table}{0}
        \setcounter{equation}{0}
        \renewcommand{\thefigure}{\Alph{appendixc}.\arabic{figure}}
        \renewcommand{\thetable}{\Alph{appendixc}.\arabic{table}}
        \renewcommand{\theappendixc}{\Alph{appendixc}}
        \renewcommand{\theequation}{\Alph{appendixc}.\arabic{equation}}
        \noindent{\bf Appendix \theappendixc #1}\par\vspace{0.4cm}}

\def\abstracts#1{{
	\centering{\begin{minipage}{30pc}\tenrm\baselineskip=12pt\noindent
	\centerline{\tenrm ABSTRACT}\vspace{0.3cm}
	\parindent=0pt #1
	\end{minipage}}\par}}

\renewenvironment{thebibliography}[1]
	{\begin{list}{\arabic{enumi}.}
	{\usecounter{enumi}\setlength{\parsep}{0pt}
\setlength{\leftmargin 1.25cm}{\rightmargin 0pt}
	 \setlength{\itemsep}{0pt} \settowidth
	{\labelwidth}{#1.}\sloppy}}{\end{list}}

\newcounter{itemlistc}
\newcounter{romanlistc}
\newcounter{alphlistc}
\newcounter{arabiclistc}

\newcommand{\fcaption}[1]{
        \refstepcounter{figure}
        \setbox\@tempboxa = \hbox{\tenrm Fig.~\thefigure. #1}
        \ifdim \wd\@tempboxa > 6in
           {\begin{center}
        \parbox{6in}{\tenrm\baselineskip=12pt Fig.~\thefigure. #1}
            \end{center}}
        \else
             {\begin{center}
             {\tenrm Fig.~\thefigure. #1}
              \end{center}}
        \fi}

\newcommand{\tcaption}[1]{
        \refstepcounter{table}
        \setbox\@tempboxa = \hbox{\tenrm Table~\thetable. #1}
        \ifdim \wd\@tempboxa > 6in
           {\begin{center}
        \parbox{6in}{\tenrm\baselineskip=12pt Table~\thetable. #1}
            \end{center}}
        \else
             {\begin{center}
             {\tenrm Table~\thetable. #1}
              \end{center}}
        \fi}

\def\@citex[#1]#2{\if@filesw\immediate\write\@auxout
	{\string\citation{#2}}\fi
\def\@citea{}\@cite{\@for\@citeb:=#2\do
	{\@citea\def\@citea{,}\@ifundefined
	{b@\@citeb}{{\bf ?}\@warning
	{Citation `\@citeb' on page \thepage \space undefined}}
	{\csname b@\@citeb\endcsname}}}{#1}}

\newif\if@cghi
\def\cite{\@cghitrue\@ifnextchar [{\@tempswatrue
	\@citex}{\@tempswafalse\@citex[]}}
\def\citelow{\@cghifalse\@ifnextchar [{\@tempswatrue
	\@citex}{\@tempswafalse\@citex[]}}
\def\@cite#1#2{{$\null^{#1}$\if@tempswa\typeout
	{IJCGA warning: optional citation argument
	ignored: `#2'} \fi}}

\def\fnt#1#2{\footnotetext{\kern-.3em
	{$^{\mbox{\sevenrm #1}}$}{#2}}}

 1
 1
 1

\font\tenbf=cmbx10
\font\tenrm=cmr10
\font\tenit=cmti10

\font\ninerm=cmr9

\begin{document}

\mbox{}\vspace*{-1cm}\hspace*{8cm}\makebox[7cm][r]{\large  HU-SEFT R
1994-18} \mbox{}\vspace*{-0cm}\hspace*{9cm}\makebox[7cm][r]{\phantom
{HU-SEFT R 1994-18}} \vspace*{0.5cm}
 \hspace*{10.9cm} \makebox[4cm]{} \vfill

\centerline{\tenbf SUPERSYMMETRIC LEFT-RIGHT MODEL AND}
\baselineskip=16pt
\centerline{\tenbf ITS PHENOMENOLOGICAL
IMPLICATIONS\footnote{Talk given by Katri Huitu at "Physics from Planck
scale to electroweak scale", Warsaw, Poland 21-24 September 1994.}}
\vspace{0.8cm}
\centerline{\tenrm KATRI HUITU, MARTTI RAIDAL}
\baselineskip=13pt
\centerline{\tenit Research Institute for High Energy Physics}
\baselineskip=12pt
\centerline{\tenit P.O.Box 9, FIN-00014 University of Helsinki, Finland}
\vspace{0.3cm}
\centerline{\tenrm and}
\vspace{0.3cm}
\centerline{\tenrm JUKKA MAALAMPI}
\baselineskip=13pt
\centerline{\tenit Department of Theoretical Physics,}
\baselineskip=12pt
\centerline{\tenit P.O.Box 9, FIN-00014 University of Helsinki, Finland}
\vspace{0.9cm}
\abstracts{We review here our study of a supersymmetric
left-right model (SLRM).
In the model the $R$-parity is spontaneously broken.
Phenomenologically novel feature of the model is the
occurrance of the doubly charged
particles in the Higgs sector, which are possibly light
enough to be seen in the next linear collider.
Detection of the doubly charged higgsinos in the next linear collider
is discussed.
}

\rm\baselineskip=14pt
\section{Introduction}
The left-right symmetric electroweak model based on the \ssu \
symmetry has many attractive features. In particular, in the see-saw
mechanism it offers a beautiful and very natural explanation for the
lightness of the ordinary neutrinos. On the other hand, like in the
Standard Model it has a hierarchy problem in the scalar sector, which
can be solved by making the theory supersymmetric.

The left-right models are especially interesting, if
the experiments on solar \cite{sun}
and atmospheric \cite{atmos} neutrinos
continue to show deviation from the standard model, as well as the
existence of the hot dark matter component \cite{COBE} explaining some features
of the power spectrum of density fluctuations of the Universe persists.
All these results seem to indicate that neutrinos indeed have a
small mass.

To achieve the see-saw mechanism, the {\ssu } symmetry
has to be broken by scalar triplets of \sur .
A novel feature of the model is that the triplet superfields contain among
others also doubly charged particles.

\vfill
\pagebreak
\section{The model}
The model is described by the superpotential
\bea
W & = & h_ {\phi Q} \widehat Q_{L}^{T}i\tau_2 \widehat \phi  \widehat Q_{R}^c
+ h_{ \chi Q} \widehat Q_{L}^{T} i\tau_2\widehat
\chi  \widehat Q_{R}^c \nonumber \\
&&+h_ {\phi L} \widehat L_{L}^{T}i\tau_2 \widehat \phi  \widehat L_{R}^c
+h_{ \chi L} \widehat L_{L}^{T} \widehat i\tau_2 \widehat\chi
 \widehat L_{R}^c
+h_{\DE} \widehat L_{R}^{cT} i\tau_2
{\widehat \DE }  \widehat L_{R}^c \nonumber\\
&& + \mu_1 {\rm Tr} (i\tau_2 \widehat \phi^T i\tau_2 \widehat \chi )
+\mu_2  \Tr (\widehat \DE \widehat \D ) ,\label{pot}
\eea
\noindent
where $\widehat Q_{L (R)}$ denote the left (right) handed quark superfield
doublets and similarly for the leptons $\widehat L_{L (R)}$.
The triplet and the bidoublet Higgs superfields of {\ssu } are given by
\bea
\label{higgses}
&&\widehat\DE =\matr{cc}\widehat\DE^-/\sqrt{2} & \widehat\DE^{0}\\
 \widehat\DE^{--}&-\widehat\DE^{-}/\sqrt{2} \ematr ,
\,\,\,\,\,\,\,\,\,\,\,\,
\widehat\delta =\matr{cc}\widehat\delta^{+}/\sqrt{2}& \widehat\delta^{++}\\
 \widehat\delta^{0} &-\widehat\delta^{+}/\sqrt{2} \ematr
\,\,\,\,\,\,\,\,\,\,\,\, \nonumber\\
&& \widehat\phi =\matr{cc}\widehat\phi_1^0&
\widehat\phi_1^+\\\widehat\phi_2^-&\widehat\phi_2^0
\ematr ,
\,\,\,\,\,\,\,\,\,\,\,\,
\widehat\chi =\matr{cc}\widehat\chi_1^0&
\widehat\chi_1^+\\\widehat\chi_2^-&\widehat\chi_2^0
\ematr  ,
\eea
 \noindent
where the different fields transform as $\widehat\DE \sim ({\bf 1,3,}-2)$,
$\widehat\delta \sim ({\bf 1,3,}2)$, $\widehat\phi \sim ({\bf 2,2,}0)$, and
$\widehat\chi \sim ({\bf 2,2,}0)$.
Corresponding to each scalar multiplet with non-zero $U(1)$
quantum number,
one has to include another multiplet with an opposite $U(1)$
quantum number in order to avoid chiral anomalies for the fermionic
superpartners.
Also another bidoublet Higgs superfield is added to get a nontrivial
Kobayashi-Maskawa matrix.

We find a region in the parameter space for which
the scalar fields in the minimum have the following
vacuum expectation values \cite{HM}:
\be
\langle\DE^0 \rangle = v_\DE,\;\; \langle\delta^0 \rangle=v_\delta,\;\;
\langle\phi_1^0 \rangle =\ka_1,\;\; \langle\chi_2^0 \rangle =\ka_2,\;\;
\langle\tilde\nu \rangle =\sigma_R.
\ee
Applying the minimization conditions $\del V/\del\ka_1=\del V/\del\ka_2=
\del V/\del v_\DE=\del V/\del v_\delta=\del V/\del \sigma_R =0$ one can find
the scalar masses.
In the minimum the $mass^2$ of all the scalars in the Higgs sector must be
positive.
This requirement has fundamental consequences for the $R$-parity,
$R=(-1)^{3(B-L)+2s}$.
The $R$-parity is automatically
conserved in Lagrangian in this type of models \cite{FIQ}, but it
may be broken spontaneously if $\langle\tilde\nu \rangle\not= 0$.
In the case of conserved $R$-parity, \ie $\langle\tilde\nu_{R,L} \rangle=0$,
the pseudoscalar mass matrix
is given by four
two by two blocks.
One of the blocks contains the sneutrinos and we need not consider
it here.
Two of the blocks contain the Goldstone bosons which make
two of the neutral gauge bosons massive.
The physical pseudoscalar particles have the masses \cite{HM}

\bea
m_{A_1}^2&=& m_{\phi \chi}^2\left(\frac {\ka_1}{\ka_2} + \frac {\ka_2}{\ka_1}
\right) ,\;\;\;\;\;\;\;
m_{A_2}^2= m_{\Delta \delta }^2\left( \frac {v_\delta }{v_\DE} +
\frac {v_\DE }{v_\delta}
\right) ,
 \nonumber \\
m_{A_{3,4}}&=& \h \left\{  m_{A_1}^2 \pm \left[  \right. m_{A_1}^4+
4(m_{W_R}^2\cos 2\gamma  -m_{W_L}^2\cos 2\beta )^2    \right.
\nonumber \\
&&  -4(m_{W_R}^2\cos 2\gamma  -m_{W_L}^2\cos 2\beta )
m_{A_1}^2 \cos 2\beta \left.\left. \right] ^{1/2} \right\} ,
\eea

\noindent
where it is defined
$\tan^2\gamma = ({ v_\delta ^2 + \h \ka_1^2})/({ v_\DE ^2 + \h \ka_2^2})$,
$\tan\beta = {\ka_2}/{\ka_1} $, and $\tan\delta
={v_\delta }/{ v_\DE } $.

On the other hand the masses of the doubly charged scalars are given by
\be
m_{H^{++}_{1,2}}^2=\h\left\{ m_{A_2}^2\pm \sqrt{m_{A_2}^4
+8 m_{W_R}^2\cos 2\gamma
[m_{A_2}^2 \cos 2 \delta + 2m_{W_R}^2 \cos 2 \gamma ]} \right\}.
\ee
\noindent
It is easily seen that trying to make both pseudoscalar and doubly charged
$mass^2$ positive one ends in contradiction.
Necessarily at least one of the $\langle\tilde\nu \rangle\not= 0$.

However, if the right-sneutrino has a vev, it is found that
ranges of parameters exist, where all the squares of the Higgs masses
are positive.
Furthermore, one of the doubly charged Higgses turns out to be
lighter than 500 GeV for $h_\DE \leq 0.8$.
The majority of the scalars are heavier than 1 TeV.

\section{Testing SLRM in the colliders}
Another particle of interest in the Higgs sector is the
supersymmetric counterpart of the doubly charged Higgs.
This particle is very suitable for experimental
search for many reasons.
It is doubly charged, which means that it
does not mix with other particles.
Consequently its mass is given by
a single parameter, the susy Higgs mixing parameter $\mu_2$.

The next generation linear electron colliders will, besides the usual
$e^+e^-$ reactions, be able to work also in $e^-e^-$, $e^-\gamma$ and
$\gamma\gamma$ modes.
The high energy photon beams can be obtained by back-scattering an intensive
laser beam on high energy electrons.
The doubly charged higgsinos can be produced in any of these operation modes
\cite{HMR1}:
\be
e^+e^-\to \tilde\Delta^{++}\tilde\Delta^{--}\label{reaction1},
\ee
\be e^-e^-\to \tilde\DE^{--}\tilde\chi^0\label{reaction2},
\ee
\be \gamma e^-\to \tilde\l^+\tilde\Delta^{--}\label{reaction3},
\ee
\be \gamma\gamma \to \tilde\Delta^{--}\tilde\Delta^{++}.\label{reaction4}
\ee
We have chosen these reactions for investigation because they   all have a
clean experimental signature: a few hard leptons and missing energy.
Furthermore, they
all have very small background from other processes.

In large regions of the parameter space, the kinematically favoured decay
mode of the triplet higgsino is
$\tilde \DE^{++}\rightarrow \tilde l^+ l^+$.
Which of the slepton decay modes is dominant, depends on kinematics, but one
possibility is the decay to a lepton and the lightest neutralino:
$\tilde l \rightarrow l\tilde \chi^0$.
The experimental signature of the doubly charged higgsino could
be then
\be
\tilde \DE^{--}\rightarrow \tilde l^- l^- \rightarrow l^- l^- \tilde \chi ^0,
\ee
where $l$ can be any of the $e,\,\mu,$ and $\tau$ with practically equal
probabilities.
The experimental signal of reactions (\ref{reaction1}) and (\ref{reaction4})
would be then four leptons and missing energy.
The total cross section of reaction (\ref{reaction1}) for the collision
energy $\sqrt{s} =$1 TeV and the
slepton and higgsino masses in the range of 100--400 GeV is about 0.5 pb.
The reaction (\ref{reaction4}) is a model independent way to
produce doubly charged higgsinos, since it depends only on the
parameter $\mu_2$.
For $\mu_2 \lsim 300$ GeV its cross section is larger than 1 pb.

The slepton pair production
\be e^+e^- \rightarrow \tilde l^+ \tilde l^- \ee
tests also the Higgs sector of the theory since the process
is mediated among others by the doubly charged higgsino \cite{HMR2}.

The selectron pair production in supersymmetric LR-model has a
larger cross section than the corresponding process
in the MSSM by about an order of
magnitude~\cite{HMR2}.
This is due to two factors, firstly the number of
gauginos in t-channel is larger and secondly the triplet higgsino contribution
in u-channel
is large, though dependent on the unknown triplet higgsino
coupling to the electron and selectron.

The u-channel
exchange of the doubly charged higgsino
occurs only for a right-handed electron and a
left-handed positron ($P_{+-}$ polarization), whereas in s- and
t-channel
processes also other chirality combinations may enter.
Use of polarized
beams  could therefore give us more information of the triplet
higgsino contribution.
For $P_{+-}$ polarization there is a peak in the backward direction
in the angular distribution of the final state electron.
Whether there is a forward peak, depends on the neutralino content.
For gaugino dominated neutralinos, the peak exists, but for higgsino
dominated neutralinos it is suppressed by the lepton Yukawa couplings.

The  cross section of the  pair production of smuons and
staus are in general expected to be smaller than that of
selectron pair production, since the neutralinos do not contribute.
On the other hand the cross sections are in general larger than in
the case of the MSSM because of the nondiagonal couplings of the
triplet higgsinos.

\vspace{0.5cm}
\noindent
The work has been supported by
the  Academy of Finland.

\end{document}